\documentclass[aps,preprint,showpacs,groupedaddress]{revtex4}  % for double-spaced preprint

\usepackage{graphicx}  % needed for figures
\usepackage{dcolumn}   % needed for some tables
\usepackage{bm}        % for math
\usepackage{amssymb}   % for math
\usepackage{float}
\begin{document}

\def\be{\begin{equation}}
\def\ee{\end{equation}}

\title{Magnetization reversal in isolated and interacting single-domain nanoparticles}
\author{H. Kesserwan}
\author{G. Manfredi}
\author{J.-Y. Bigot}
\author{P.-A. Hervieux}
\affiliation{Institut de Physique et Chimie des Mat$\acute{e}$riaux de Strasbourg, UMR 7504, CNRS and Universit$\acute{e}$ de Strasbourg, 23 rue du Loess, F-67034 Strasbourg, France}
\vskip 0.25cm

\date{\today}

\begin{abstract}
Computational and experimental results on the thermally-induced magnetization reversal in single-domain magnetic nanoparticles are reported. The simulations are based on the direct integration of the Fokker-Planck equation that governs the dynamics of the magnetic moment associated with the nanoparticles. A mean field approximation is used to account for the influence of the dipolar interaction between nanoparticles. It is shown that the interactions can either speed up or slow down the reversal process, depending on the angle between the external magnetic field and the axis of easy magnetization. The numerical results are in good agreement with experimental measurements on cobalt-platinum nanoparticles.
\end{abstract}

\pacs{75.20.-g; 75.60.Jk; 75.75.Jn}
\maketitle

%\section{\label{sec:level1}First-level heading}
% sections are not used for PRL papers

\section{Introduction} \label{sec:intro}
Single-domain magnetic nanoparticles constitute an attractive system for fundamental research as well as for advanced technological applications. The use of single-domain magnetic nanoparticles is expected to increase the data storage density to several petabit/inch$^{2}$ (10$^{15}\rm cm^{-2})$ \cite{weller,Gubin} in the near future. However, when the size of the nanoparticles is reduced, the superparamagnetic regime can be attained and the magnetization fluctuates under the effect of thermal excitations \cite{weller1}. This effect is a major drawback for technological developments and it is therefore important to investigate and understand the thermally-induced magnetization reversal in these systems.

In isolated single-domain magnetic nanoparticles, the magnetization reversal by thermal activation is well described by the N\'{e}el-Brown model \cite{neel,brown}. According to this model, the thermal fluctuations cause the magnetic moment to undergo a Brownian-like motion about the axis of easy magnetization, with a finite probability to flip from one equilibrium direction to another.
From an energetic point of view, the two minima associated to the equilibrium positions are separated by a barrier due to the magneto-crystalline and shape anisotropies. The corresponding Arrhenius-type superparamagnetic relaxation time can be written as
$ \tau= \tau_0 \exp(\Delta E/k_{B}T)$, where $\Delta E$ is the energy barrier between the two easy directions of the magnetization, $k_B$ is the Boltzmann constant, and $T$ is the temperature.
The typical time $\tau_{0}$ is not well known experimentally and is estimated to be of the order $10^{-10}-10^{-12}$s for magnetic nanoparticles.

The Arrhenius exponential law describes well the behavior of isolated (i.e., non-interacting) nanoparticles. The effect of dipole-dipole interactions on the relaxation time and, more generally, on the reversal process has been studied in several works, both theoretical \cite{Hansen,Otero} and experimental \cite{Morup,Dormann, Allia, Poddar}. In spite of significant progress, it is still a controversial issue, as opposite dynamical switching behaviours have been reported in the literature \cite{Hansen}. This is of course a major issue for the development of smaller and faster switching memory devices.

In this work, we present numerical calculations of the relaxation times for the magnetization reversal in isolated and interacting single-domain ferromagnetic nanoparticles with uniaxial anisotropy. Further, in order to validate our theoretical approach, we have studied experimentally the dynamics of close packed cobalt-platinum core-shell nanoparticles which can be considered as a model system owing to their spherical shape and small size dispersion (less than 5\%). The relaxation time is determined by performing magneto-optical measurements using femtosecond laser pulses. As shown in the following sections, our theoretical model is in good agreement with the experimental results.

\section{Model} \label{sec:model}
Micromagnetic simulations of thermally activated magnetic systems are generally performed using either the Langevin Dynamics (LD) \cite{chantrell}, which is based on the direct integration of the stochastic Landau-Lifshitz-Gilbert equation \cite{brown}, or the so-called `time-quantified' Monte Carlo method (TQMC) \cite{nowak}, which is a generalization of standard Monte Carlo techniques to the time-dependent regime.
The TQMC method was recently extended to the case of correlated particles with nearest-neighbors exchange interactions \cite{Cheng}, although this approach was limited to small systems ($10 \times 10$ square lattice) with short-range interactions.

Adapting the above methods to the case of long-range dipolar interactions would require prohibitive computational times and memory storage, particularly when the number of nanoparticles is large. In order to circumvent this problem, we have developed an approach based on the direct integration of the Brown-Fokker-Planck equation \cite{brown}, which describes the time evolution of the probability distribution $W(\theta,\phi,t)$ of the magnetic moment of a nanoparticle. $\theta$ and $\phi$ are the polar angles determining the orientation of the magnetization vector $\bf{m}$ with respect to the axis of quantification $z$ ($0\leq\theta\leq\pi$) and its projection on the equatorial plane ($0\leq\phi\leq 2\pi$).
In the following, we consider the case of axial symmetry, for which all quantities do not depend on $\phi$. Dipole-dipole interactions are treated in the framework of the mean field approximation.

For an isolated nanoparticle in an external magnetic field $\bf{H}_{0}$ aligned along the axis of easy magnetization, the Brown-Fokker-Planck equation reads as \cite{brown}:
\begin{equation}\label{eq:fokker}
    \frac{\partial W}{\partial t}=\frac{1}{\sin\theta}\frac{\partial}{\partial \theta}\left[\sin\theta\Big(h'\frac{\partial E}{\partial \theta}W+k'\frac{\partial W}{\partial \theta}\Big)\right],
\end{equation}
\be
h'=\frac{k'}{k_B T}=\frac{\alpha\gamma_{0}}{(1+\alpha^{2})VM_{s}},
\ee
where $V$ is the volume of the particle, $\alpha$ is the Gilbert damping constant, $\gamma_{0}$=1.76$\times$10$^{11} ~\rm T^{-1}s^{-1}$ is the gyromagnetic factor, and $M_{s}$ is the magnetization at saturation. The total energy of a nanoparticle is given by $E(\theta)=KV\sin^{2}\theta-\mu_{0}\bf{m}\cdot\bf{H}_{0}$, where $K$ is the anisotropy constant and $\mu_0$ is the vacuum magnetic permeability.

It is useful to rewrite Eq. (\ref{eq:fokker}) using dimensionless quantities, by introducing the normalized time variable $\widehat{t} = t~\frac{\alpha\gamma_{0}k_{B}T}{(1+\alpha^{2})VM_{s}}$ and defining
$x=\cos\theta$, with $-1\le x\le 1$. The resulting equation can be put in the form of an advection-diffusion equation:
\begin{equation}\label{eq:fokker-norm}
    \frac{\partial W}{\partial \widehat{t}}=\frac{\partial }{\partial x}\Big(U(x)W\Big)+\frac{\partial }{\partial x}\left(D(x)\frac{\partial W}{\partial x}\right),
\end{equation}
where $D(x) = 1-x^2$, $U(x)=(x^2-1)(Ax+B)$, and $A$ and $B$ are two dimensionless constants representing respectively the anisotropy energy and the energy due to the external field:
\be
  A= \frac{2KV}{k_{B}T}~; ~~~~
  B= \frac{\mu_{0}VM_{s}H_{0}}{k_{B}T}.
\ee
\begin{figure}[htb]
% \centerline{\includegraphics[width=6cm, height=4.5cm]
 \centerline{\includegraphics[width=8cm, height=6cm]
  {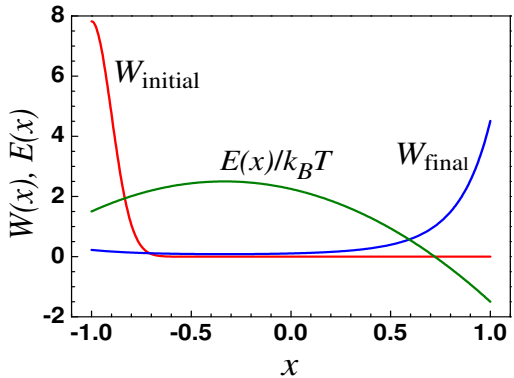}}
  \caption{(color online). A sketch of the initial (red line) and final (blue line) distributions of the moments with respect to the energy barrier (green line).}\label{figure_1}
\end{figure}

We solved Eq. (\ref{eq:fokker-norm}) using a finite difference technique with boundary conditions chosen so that the total probability $\int_{-1}^{1}W(x,t)dx$ remains constant in time.
Initially, the magnetic moments are set near the local equilibrium with higher energy, which is situated at $x=-1$ ($\theta=\pi$), so that the probability distribution $W(x,0)$ is given by a half-Gaussian with maximum at $x=-1$ (red curve on Fig. \ref{figure_1}). Physically, this corresponds to a metaequilibrium with a magnetic moment oriented in the direction opposite to the external magnetic field. Our procedure consists in computing the average time necessary for the magnetic moment to reverse its orientation and align with the external field. This configuration minimizes the energy for all angles $\theta$ and therefore constitutes an accurate description of the equilibrium, given by the Boltzmann law $W_{\rm final}={\rm const.} \times e^{-E/k_B T}$ (blue curve on Fig. \ref{figure_1}).

In order to obtain the relaxation time, we compute the integral $I_{-}(t)=\int_{-1}^{0}W(x,t)dx$, which represents the probability of finding the magnetic moment on the lower part of the equatorial plane, and then we fit this quantity with a decaying exponential. This procedure is repeated for different values of the temperature. For comparison with experimental results, we will apply our numerical simulations to the case of cobalt nanoparticles, which are ferromagnetic at low temperature (typically, the superparamagnetic blocking temperature is $\sim 100$ K for nanoparticles with a diameter less than 10 nm).

\begin{figure}[H]
%  \centerline{\includegraphics[width=6cm, height=4.5cm]
  \centerline{\includegraphics[width=8cm, height=6cm]
  {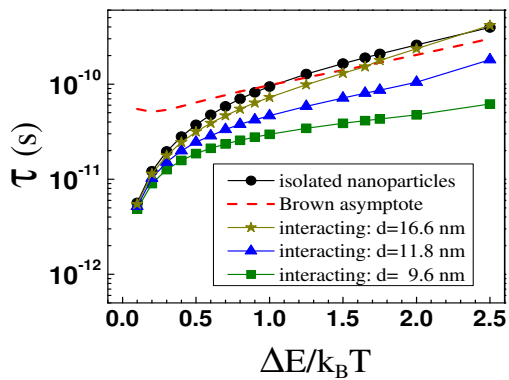}}
  \setlength{\abovecaptionskip}{0 pt}\caption{(color online). Relaxation times for isolated nanoparticles (circles) and interacting nanoparticles for different interparticle distances: $d=9.6$ nm (squares), $d=11.8$ nm (triangles), and $d=16.6$ nm (stars). The occupation probability is $p=0.5$. The damping constant $\alpha=1$.}\label{figure_2}
\end{figure}

\section{Numerical results} \label{sec:results}
First, we focus on the case of non-interacting (isolated) particles. We considered spherical Co nanoparticles with a diameter of 9.5 nm, an anisotropy constant $K=4.2\times10^{5}\rm ~Jm^{-3}$, and a saturation magnetic moment $m_{s}=VM_{s}$=6.45$\times 10^{-19}\rm ~JT^{-1}$. The external magnetic field is $H_{0}=1.59\times 10^{5}\rm ~Am^{-1}$.
%
%These values were chosen because they correspond to the experimental results
%discussed later (see Sec. \ref{sec:exp}).

We computed the relaxation time $\tau$ as a function of the density barrier $\Delta E = KV(1-\mu_0 H_0 M_s/2K)^2$. The results for isolated nanoparticles are plotted in Fig. \ref{figure_2} (black circles) and are in agreement with Brown's expression (dashed line) at low temperatures, with a $10 \%$ accuracy. At higher temperatures, the numerical solution departs from Brown's expression and the computed relaxation time tends to zero. These results are consistent with those obtained using different numerical techniques such as time-quantified Monte Carlo \cite{nowak} and thus fully validate our approach.

Let us now focus on the more challenging problem of magnetic nanoparticles interacting via the magnetic dipolar interaction. The case of two particles was considered by Rod\'{e} et al. \cite{rod}. The generalization to many-particle systems is far from trivial \cite{Hansen,Otero}, as the computational complexity increases as the square of the number of particles $N^2$. This can be reduced to $N \ln N$ by making use of fast Fourier transforms \cite{yuan,Hinzke}, but still requires considerable computing time for large systems.

Here, we use an alternative approach based on the mean field approximation \cite{Denisov1,Denisov2,Denisov3}, whereby the effect of the dipolar interaction on one particle exerted by all the others is expressed as a self-consistent magnetic field. This approximation neglects higher-order fluctuating correlations between the magnetic moments, which are a further source of random Brownian motion beyond the usual fluctuations induced by thermal effects. In the regimes considered here, which operate at relatively high temperatures, the thermal fluctuations are likely to be dominant so that the mean field approximation should be accurate enough.

We consider a system of interacting nanoparticles distributed over a spatial lattice. At sufficiently low temperature (or high energy barrier), the magnetic moments fluctuate in the vicinity of the positive and negative directions of the $z$ axis. In this case, the self-consistent field is also aligned along the $z$ direction and can be written as ${\bf H}_{D}= H_D(t){\bf e}_{z} = 8pSd^{-3}m_{z}(t){\bf e}_{z}$ , where $S$ is a constant that depends on the geometry of the lattice, $0\le p\le 1$ is the probability of occupation of the sites, $d$ is the center-to-center interparticle distance, and $m_{z}(t)$ is the average $z$ component of the total magnetic moment of the system:
\be
m_{z}(t)=m_{s}\int_{0}^{\pi}\cos\theta W(\theta,t)d(\cos\theta).
\label{eq:mz}
\ee
For a two-dimensional lattice \cite{Denisov1}, the mean field model yields a value $S\approx -1.129$, where the negative sign signals that the dipolar interaction is antiferromagnetic.

In the mean field approximation, the nanoparticles are considered to be independent, with all particles experiencing the effective dipolar mean field ${\bf H}_{D}(t)$ as an external magnetic field. Thus, the total energy becomes $E(\theta)=KV\sin^{2}\theta-\mu_{0}\bf{m}\cdot (\bf{H}_{0}+
{\bf H}_{\it D})$. This entails a modification of $U(x)$ in Eq. (\ref{eq:fokker-norm}), which becomes $U(x,t)= (x^2-1)[Ax+B+C(t)]$, where $C(t)$ is a term describing the strength of the dipolar interaction:
\be
C(t)=\frac{\mu_{0}VM_{s}}{2\pi k_{B}T}~H_D(t).
\label{eq:ct}
\ee
We emphasize that, within this approach, the relevant Fokker-Planck equation (\ref{eq:fokker-norm}) becomes nonlinear, because the dipolar term depends on $W$ through Eq. (\ref{eq:mz}).

The numerical results for interacting particles are shown in Fig. \ref{figure_2}, for the case with $p=0.5$ and three values of the interparticle distance $d$. The Gilbert damping constant is taken to be $\alpha=1$, a value that is consistent with recent measurements on small Co nanoparticles \cite{Andrade}. It is found that the relaxation time decreases with decreasing interparticle distance, in agreement with previous experimental measurements \cite{Morup}. Beyond a certain distance ($d \approx 16$ nm in the case of Fig. \ref{figure_2}), the dipolar interaction becomes negligible and the result for the corresponding isolated nanoparticles is retrieved. In contrast, the relaxation time decreases with increasing site occupation probability $p$. In summary, the magnetization reversal process is accelerated for short interparticle distances and large concentrations.

In all the preceding results, the external magnetic field ${\bf H}_{0}$ was set parallel to the axis of easy magnetization. When the field makes an angle $\Psi$ with respect to the easy axis, the energy profile becomes:
\be
E(\theta)=KV\sin^{2}\theta-\mu_{0}m_{s} H_{0}\cos(\theta-\Psi)-\mu_{0}{\bf m}\cdot{\bf H}_{D}.
\ee
Coffey et al. \cite{Coffey} found that the relaxation time in isolated nanoparticles varies with the angle and has a minimum at $\Psi=\pi/4$ and a maximum at $\Psi=\pi/2$. Our mean field model \cite{Denisov2} assumes that the self-consistent dipolar field is always aligned along the axis of easy magnetization, which is a reasonable assumption only when the external field is either parallel ($\Psi=0$) or antiparallel ($\Psi=\pi$) to $z$.
\begin{figure}[H]
%  \centerline{\includegraphics[width=6cm, height=4.5cm]
  \centerline{\includegraphics[width=8cm, height=6cm]
  {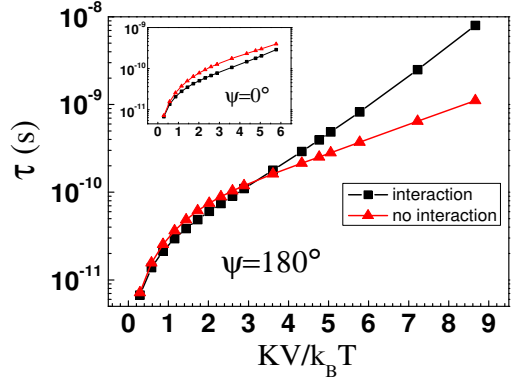}}
  \setlength{\abovecaptionskip}{0 pt}\caption{(color online). Relaxation time $\tau$ vs the normalized inverse temperature, for an external field antiparallel to the direction of easy magnetization ($\Psi=180^\circ$). The inset shows the same result for $\Psi=0^\circ$.}\label{figure_3}
\end{figure}

We have studied the relaxation time for nanoparticles with the following physical parameters:  $K=4.2\times 10^{5} {\rm Jm}^{-3}$, $m_{s}= 8\times 10^{-20} {\rm JT}^{-1}$, $V=5.6\times 10^{-26} \rm m^3$, $d=8.6\times 10^{-9} \rm m$, $S=-1.129$ and $p=1$. The results are shown in Fig. \ref{figure_3} for different values of the lattice temperature.
When the external field is parallel to the $z$ axis ($\Psi=0$), the effect of the dipolar interactions is to shorten the relaxation time, i.e., to accelerate the reversal of the magnetization. In contrast, for the antiparallel configuration ($\Psi=\pi$), the relaxation time is {\em longer} when the dipolar interactions are taken into account, at least for temperatures below a certain threshold (here the threshold is situated near $KV/k_B T=3$, corresponding to a temperature $T=567$ K). This effect can be understood by noticing that the dipolar field either adds or subtracts to the external magnetic field, depending on the orientation of the latter. Nevertheless, since the dipolar field -- and thus the energy barrier -- depends self-consistently on $W(\theta,t)$, nonlinear effects might play an important role in the reversal dynamics. Further work will be necessary to clarify these issues.

\section{Comparison to experiments} \label{sec:exp}
In order to compare our theoretical results to experiments,
we have measured the relaxation times in core-shell cobalt-platinum nanoparticles.
The CoPt core-shell nanoparticles are made by a redox transmetalation reaction. They have a spherical shape made of a cobalt core with an average diameter of 5 nm and a platinum shell with average thickness 1.5 nm as measured by high resolution electron microscopy (JEOL 2100F microscope). The size dispersion is less than 5\%.  These nanoparticles are assembled into a compact bulk pellet by cold pressing under 160 Pa. Without any thermal process, the nanoparticles display a superparamagnetic behavior at room temperature with a blocking temperature of $66 \pm 3$ K as measured by Zero Field Cooling/ Field Cooling in a SQUID apparatus. At 5 K, the nanoparticles are ferromagnetic with a coercive field $H_c = 7.56\times 10^3~\rm A m^{-1}$ and a magnetization at saturation $M_s = 4 \times 10^4~ \rm A m^{-1}$. The inter-particle distance (Co-Co) is estimated to be 8.6 nm and the particle concentration is large
enough so that in the simulations we can safely assume $p = 1$.

To measure the relaxation times in core-shell cobalt-platinum nanoparticles we have performed time-resolved pump-probe magneto-optical Kerr measurements. The probe pulses are obtained from an amplified Ti:Sapphire laser functioning at a wavelength   $\lambda = 800$ nm and with a duration of 120 fs. The pump pulses with a wavelength $\lambda = 400$ nm are obtained by frequency doubling in a $\beta-\rm BaB2O4$ nonlinear crystal and have a duration of 150 fs. Upon heating the nanoparticles, the pump pulses also trigger a motion of precession of the magnetization that can directly be observed on the time resolved magneto-optical Kerr signals. The experimental times $\tau$ are obtained by fitting the relaxation of this laser-induced precession, and are determined for several values of an external static magnetic field (which is parallel to the direction of easy magnetization, i.e., $\Psi=0$).

It must also be stressed that the laser pulse penetrates into the sample for a distance of the order of $10-15$ nm, corresponding to just a few layers of nanoparticles. This penetration depth is much smaller than the sample dimensions, so that the excited region is approximately two-dimensional, in accordance with our theoretical assumptions.

\begin{figure}[H]
% \centerline{\includegraphics[width=6cm, height=4.5cm]
  \centerline{\includegraphics[width=6cm, height=8cm, angle=-90]
  {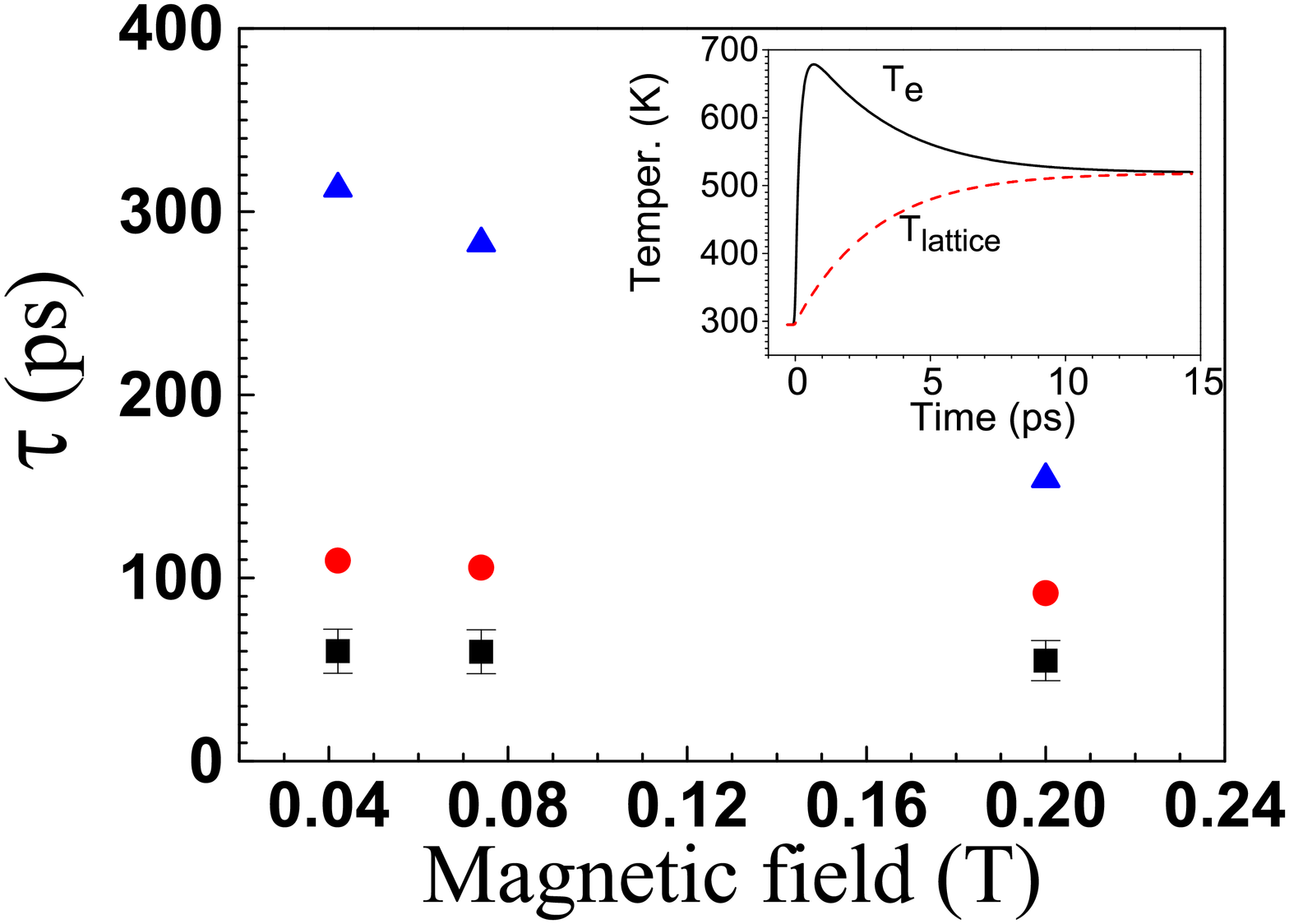}}
  \setlength{\abovecaptionskip}{0 pt}\caption{(color online). Relaxation times for three values of the external magnetic field: experimental results (black squares); simulations including interactions (red circles); simulations without interactions (blue triangles). The simulations were carried out at a temperature $T=520$ K. The inset shows the evolution of the electron (black solid line) and lattice (red dashed line) temperatures following laser excitation, obtained with a two-temperature model.}\label{figure_4}
\end{figure}

The important physical parameter that has to be determined is the lattice temperature. The nanoparticles are excited by a short laser pulse that heats up the electron gas, which subsequently cools down by exchanging energy with the lattice. After a few picoseconds, the electrons and the lattice are in thermal equilibrium at a temperature that is usually a few hundred degrees above the initial temperature. This equilibrium temperature can be estimated using a simple two-temperature model \cite{bigot, beaurepaire}, which for parameters relevant to the present experiment yields an equilibrium temperature around $T=520$ K (inset of Fig. \ref{figure_4}).

This value of the temperature was used in our Fokker-Planck simulations of the magnetization dynamics. The results are plotted in Fig. \ref{figure_4} and show a good agreement between the experiments and the simulations that include the dipole-dipole interactions. It must be pointed out that the experimental relaxation times are subject to large errors due to the difficulty of determining precisely the Gilbert damping, so that an agreement between theory and experiment within a factor of two is already remarkable. Importantly, the simulations {\em without} interactions are clearly in much poorer agreement, for the absolute values as well as the general trend.

\section{Conclusion} \label{sec:conclusion}
We developed a Fokker-Planck model that allows to simulate the magnetization dynamics in isolated and interacting single-domain magnetic nanoparticles. Isolated nanoparticles follow the N\'{e}el-Brown Arrhenius-like law at low temperatures (or, equivalently, large values of the energy barrier), but deviate from it at high enough temperatures. In the case of interacting nanoparticles, we made use of a mean-field approach, whereby the magnetic moment of each particle interacts with the mean dipolar field generated by all the others. Significant deviations from the Arrhenius law were observed, leading to a faster reversal process in the presence of the magnetic dipolar interaction. The simulation results were in good agreement with experimental measurements performed on Co-Pt nanoparticles. Finally, we showed that, when the external field is antiparallel to the direction of easy magnetization, the effect of the dipolar field can be to slow down, rather than speed up, the reversal process.

\end{document}